\documentclass[lettersize,journal]{IEEEtran}
\usepackage{amsmath,amsfonts}
\usepackage{algorithmic}
\usepackage{algorithm}
\usepackage{array}
\usepackage[caption=false,font=normalsize,labelfont=sf,textfont=sf]{subfig}
\usepackage{textcomp}
\usepackage{stfloats}
\usepackage{url}
\usepackage{verbatim}
\usepackage{graphicx}
\usepackage{cite}
\usepackage{amsmath,amssymb}
\usepackage{physics}
\usepackage{graphicx}
\usepackage{bm} 
\usepackage{upgreek}

\usepackage[%dvipdfm,
CJKbookmarks=false,
colorlinks,
breaklinks=true,
linkcolor=blue,
anchorcolor=blue,
citecolor=blue,
urlcolor=blue,
]{hyperref}

\begin{document}

\title{Calibration-free Rydberg Atomic Receiver for Sub-MHz Wireless Communications and Sensing}

\author{Minze Chen, Tianqi Mao,~\IEEEmembership{Member,~IEEE}, Wei Xiao, Zhonghuai Wu, Dapeng Li,~\IEEEmembership{Member,~IEEE}, Mingyao Cui, Qunsong Zeng, Dezhi Zheng, Kaibin Huang,~\IEEEmembership{Fellow,~IEEE}, and Zhaocheng Wang,~\IEEEmembership{Fellow,~IEEE} 
        % <-this % stops a space
\thanks{Minze Chen, Tianqi Mao, Wei Xiao, Zhonghuai Wu, Dapeng Li and Dezhi Zheng are with State Key Laboratory of Environment Characteristics and Effects for Near-space, Beijing Institute of Technology. Mingyao Cui, Qunsong Zeng and Kaibin Huang are with the Department of Electrical and Electronic Engineer
ing, The University of Hong Kong. Zhaocheng Wang is with Tsinghua University. The corresponding authors are Tianqi Mao and Dezhi Zheng.}%
\vspace*{-5mm}
} %



\maketitle

\begin{abstract}

The exploitation of sub-MHz (\textless 1 MHz) can be beneficial for a plethora of applications like underwater vehicular communication, subsurface exploration, low-frequency navigation etc. The traditional electrical receivers in this band are either hundreds of meters long or, when miniaturized, inefficient and bandwidth-limited, making them inapplicable for practical underwater implementations. Such obstacles can be circumvented by the emerging Rydberg atomic receiving technology, which is capable of detecting fields from DC up to the terahertz regime with compact structure. Against this background, we propose a method to detect sub-MHz electric fields without further calibration. Specifically, a physics-based model of the combined DC and AC-Stark response is established. Based on the model, we modulate the DC-Stark spectrum with the received signal and extract its amplitude by fitting the cycle-averaged, symmetric Stark-split peaks. Then we map this swing directly to the intrinsic atomic polarizability. By such operations, the proposed method can remove the dependence on electrode spacing or field-amplitude references. For performance evaluation, six-level Lindblad simulations and experiments are conducted at a low-frequency field of 30 kHz demonstrate a minimum detectable field of 5.3 \text{mV}/\text{cm}, with stable readout across practical optical-power variations. The approach manages to expand operating range of Rydberg atomic receivers below 1 MHz, and enables compact, calibration-free quantum front ends for underwater and subsurface receivers. 

\end{abstract}

\vspace{-3mm}

\section{Introduction}
\IEEEPARstart{S}{ub}-MHz band is fundamental to specialized communication and sensing applications \cite{Mao_WCM_2022}. Specifically, thanks to the excellent penetration and long-range transmission capabilities, such waves are crucial for underwater vehicular communication, subsurface exploration, and emergency communications. Traditional front ends for frequencies below 1 MHz are fundamentally size-limited by the Chu–Harrington bound, which forces efficient antennas to occupy a substantial fraction of the wavelength \cite{7842558,10234603}. For instance, trailing antennas in underwater vehicles like submarines are often several hundred meters in length for VLF communications \cite{palmeiro2011underwater,1092211}. Furthermore, when simultaneously employing different communication/sensing bands for diverse missions, the receiving platform has to maintain multiple sets of electronic devices with large-aperture antennas, which exacerbates the payload burdens. 

Against this background, recent researches focus on Rydberg atomic receivers for their wavelength independence \cite{9905692,chen2025new}, requiring only centimeter-scale atomic vapor cells as front ends to receive electric field signals. Highly excited Rydberg atoms exhibit large polarizabilities and dipole moments, and have been demonstrated to be highly sensitive and broadband \cite{gallagher1994rydberg, cui2025rydberg}. The Rydberg atomic receiver is typically dependent on the Autler-Townes (AT) splitting in electromagnetically induced transparency (EIT) spectra \cite{sedlacek2012microwave}, which is capable of accurate electric-field detection by measuring the splitting interval, proportional to the Rabi frequency. However, classical EIT-AT methods are impractical below 100 MHz due to the lack of suitable dipole-allowed transitions, diminished dipole moments and significant spectral broadening at very high principal quantum numbers \cite{li2023super}. 

% 对于低于EIT-AT的频率范围，
% 因此，
% 里德堡原子低频测量方法原理 --> 由于原子寿命只能below 1MHz  -->  需要预先标定 --> 光功率鲁棒性很差 

To circumvent this issue, a common alternative is nonresonant AC-Stark sensing \cite{fan2015atom,li2021power}. By fitting the EIT resonance, the Stark shift can be extracted with sub-percent uncertainty \cite{holloway2022electromagnetically}. 
However, this approach ceases to be valid in the sub-MHz regime. When the signal frequency falls below the EIT coherence bandwidth (typically 1 MHz) and the optical readout chain is sufficiently wideband, the atomic polarization and optical readout track the instantaneous Stark detuning \cite{delone1999ac}. The measured spectrum is therefore not a rigidly shifted EIT line but a cycle-averaged convolution of the line with the Stark modulation produced by the applied sub-MHz electric fields. Under this regime, two sub-MHz readouts are commonly used. One line of work infers the field from the edge-enhanced shoulders of the cycle-averaged EIT profile. This route is set by atomic polarizability and is therefore calibration-free, yet it exhibits low sensitivity and systematic bias because a pronounced central peak remains near the unshifted resonance and the nonuniform temporal weighting around turning points shifts the apparent extrema. Specifically, for a typical 10 MHz full width at half maximum (FWHM), fields on the order of 60 mV/cm are needed before Stark-split subpeaks separate cleanly, and cycle averaging further degrades sensitivity \cite{ma2022measurement}. A second line of work measures the time-domain EIT oscillation produced by the low-frequency field \cite{jau2020vapor}. Variants include atom-based superheterodyne detection \cite{yang2024radio} and the use of sublevels with higher field-induced polarizability to boost response \cite{li2023super,Lei:24}. These methods can reach higher sensitivity \cite{2jpt-6313}. However, the finite Rydberg state lifetime hinders faithful atomic sensing to fields below 1 MHz. Beyond this frequency, the EIT modulation averages out and the beat signal disappears. More importantly, converting measured oscillation amplitude to absolute field amplitude in these schemes still requires external calibration (such as AC-Stark references or electrode-field calculations), making the result vulnerable to laser-power fluctuations. Therefore, a calibration-free, interference-resistant technique for sub-MHz field measurement remains an open challenge that demands further investigation.

In this work, we propose a calibration-free receiving approach for sub-MHz fields. The contributions are threefold:

\begin{enumerate}
\item We develop a system model that captures the quantum dynamics under simultaneous DC- and AC-Stark effects and we analyze the mechanism underlying the modulation-averaged EIT spectrum for electric-field frequencies below 1 MHz.
\item We propose a two-step measurement that performs cycle-averaged data acquisition followed by spectral fitting of the Stark-modulated EIT line to retrieve the field without external amplitude calibration.
\item We validate the approach through matched simulations and cesium-vapor experiments under an applied 30 kHz electric field, demonstrating a minimum detectable field of 5.3 mV/cm, with stable readout under practical probe/coupling power variations.
\end{enumerate}

% 写到优势，应继续写本文的工作

\vspace{-2mm}
\section{System Model}
\label{sys}

\begin{figure}[t!]
    \centering
    \includegraphics[width=1\linewidth]{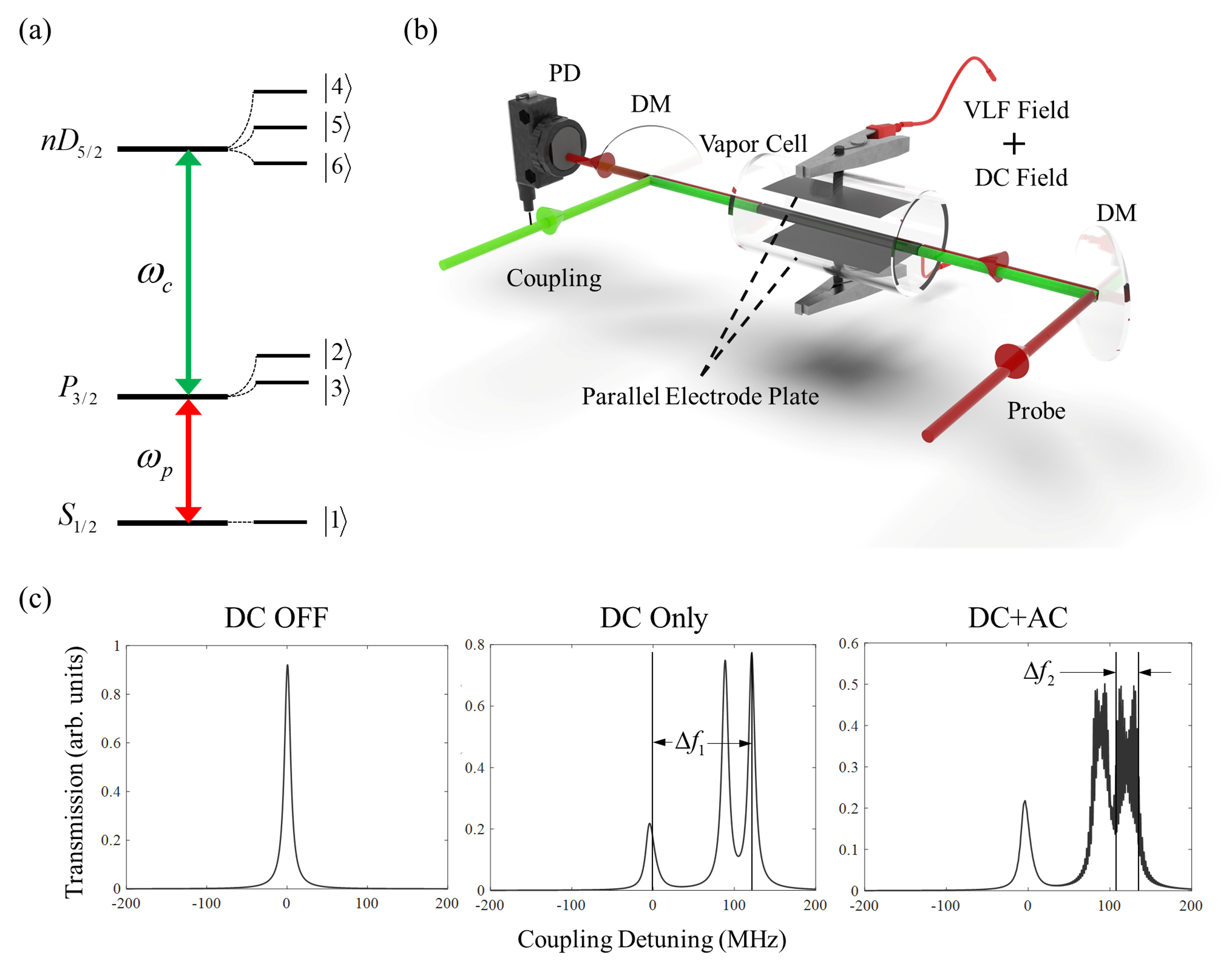} 
    \vspace{-5mm}
    \caption{(a) Energy level system and (b) experimental Setup. PD: Photodiode, DM: Dichroic Mirror. (c) Principle of the calibration-free measurement. }
    \label{fig:framework} 
    \vspace{-3mm}
\end{figure}

The degenerate three-level system undergoes splitting under an external electric field. Sublevels corresponding to different magnetic quantum numbers ($m_J$) undergo distinct Stark shifts due to their varying energy perturbations. As a result, we consider a six-level atomic cascade system based on cesium ($^{133}\mathrm{Cs}$), involving magnetic sublevels of the ground state $6\mathrm{S}_{1/2}$, intermediate state $6\mathrm{P}_{3/2}$, and Rydberg state $60\mathrm{D}_{5/2}$. Specifically, the atomic states are $|1\rangle=|6\mathrm{S}_{1/2}, m_J=\pm 1/2\rangle$, $|2\rangle=|6\mathrm{P}_{3/2}, m_J=\pm 1/2\rangle$, $|3\rangle=|6\mathrm{P}_{3/2}, m_J=\pm 3/2\rangle$, $|4\rangle=|60\mathrm{D}_{5/2}, m_J=\pm 1/2\rangle$, $|5\rangle=|60\mathrm{D}_{5/2}, m_J=\pm 3/2\rangle$, and $|6\rangle=|60\mathrm{D}_{5/2}, m_J=\pm 5/2\rangle$ as shown in Fig. \ref{fig:framework}(a). 

Atoms interact with two laser fields: a probe laser coupling ground state $|1\rangle$ to intermediate states ($|2\rangle, |3\rangle$), and a coupling laser driving transitions from intermediate to Rydberg states ($|4\rangle, |5\rangle, |6\rangle$). This two-photon excitation generates EIT resonances, strongly modulated by external electric fields.

The applied electric field consists of static and oscillating components and is given by 

\begin{equation}
E(t) = E_\mathrm{DC} + E_\mathrm{AC} \cos(2\pi f_\mathrm{AC} t),
\label{eq:Et}
\end{equation}
where $E_{\mathrm{DC}}$, $E_{\mathrm{AC}}$, and $f_{\mathrm{AC}}$ represent static amplitude, oscillating amplitude, and AC frequency. The resulting quadratic Stark shift for state $i$ is $-1/2\alpha_i E^2(t)$, with $\alpha_i$ the effective polarizability.

Under the rotating wave approximation (RWA), the Hamiltonian is given by

\begin{align}
\mathbf{H}(t) ={}& \sum_{i=1}^{6} \left[-\frac{1}{2}\alpha_i E^2(t) - \hbar\delta_i(t)\right] \ket{i}\bra{i} \nonumber\\
& + \frac{\hbar}{2} \sum_{i<j} \left( \Omega_{ij} \ket{i}\bra{j} + \Omega_{ij}^{*} \ket{j}\bra{i} \right),
\end{align}
where \(\alpha_i\) denotes the effective polarizability of the \(i\)-th atomic state. Here, the diagonal terms incorporate Stark shifts due to the electric field intensity \(E(t)\). The laser detunings \(\delta_i(t)\) characterize the difference between the atomic resonances and instantaneous laser frequencies, which in practice are typically chirped or swept across resonance conditions. The off-diagonal elements of the Hamiltonian represent coherent coupling between atomic states induced by the laser fields, with coupling strengths quantified by Rabi frequencies \(\Omega_{ij}\).

The time evolution of the atomic density matrix \(\boldsymbol{\rho}(t)\) is governed by the Lindblad master equation, a standard formalism for describing open quantum systems. This equation comprehensively captures both coherent evolution under the Hamiltonian and dissipative processes via Lindblad operators \(\mathbf{L}_k\) which is expressed by

\begin{align}
\frac{\mathrm{d}\boldsymbol{\rho}(t)}{\mathrm{d}t} ={}& -\frac{i}{\hbar}[\mathbf{H}(t), \boldsymbol{\rho}(t)] \nonumber\\
& + \sum_k \left( \mathbf{L}_k \boldsymbol{\rho}(t) \mathbf{L}_k^{\dagger}
- \frac{1}{2} \left\{ \mathbf{L}_k^{\dagger} \mathbf{L}_k, \boldsymbol{\rho}(t) \right\} \right),
\end{align}
where each Lindblad operator \(\mathbf{L}_k = \sqrt{\gamma_{ij}}\,\ket{j}\bra{i}\) corresponds to a specific decay channel $k$ from \(\ket{i}\) to \(\ket{j}\), and \(\gamma_{ij}\) is the corresponding spontaneous decay rate.

\vspace{-2mm}
\section{Proposed Measurement Method}
\label{sec:method}

The observable of primary interest in our model is the induced atomic coherence between the ground state and intermediate excited state, specifically quantified by the imaginary part of the off-diagonal density matrix elements \(\mathrm{Im}\big[\rho_{12}(t)+\rho_{13}(t)\big]\). Physically, this coherence directly relates to the probe field absorption, thus determining the transparency window and corresponding spectral features observed experimentally. 

Although the density matrix formalism fully describes atomic dynamics, an analytic steady-state line shape simplifies the electric-field retrieval process. For probe durations much longer than the period of the applied AC field, the cycle-averaged Lorentz profile is formulated as

\begin{equation}
\bar{S}(x)=\frac{1}{2\pi}\int_{0}^{2\pi}
\frac{c^{2}}
      {\bigl[x-\bigl(b+\Delta\sin\phi\bigr)\bigr]^{2}+c^{2}}\,
      \mathrm{d}\phi,
\label{eq:modulated_averaged_Lorentz}
\end{equation}
where $x$ is the probe detuning. $c$ is the Lorentz half width at half maximum (HWHM). $\phi$ denotes the instantaneous phase of the AC field during one modulation cycle. $\Delta$ is the AC-induced frequency swing, and $b$ is the unmodulated line center.

\begin{figure}[t!]
    \centering
    \includegraphics[width=1\linewidth]{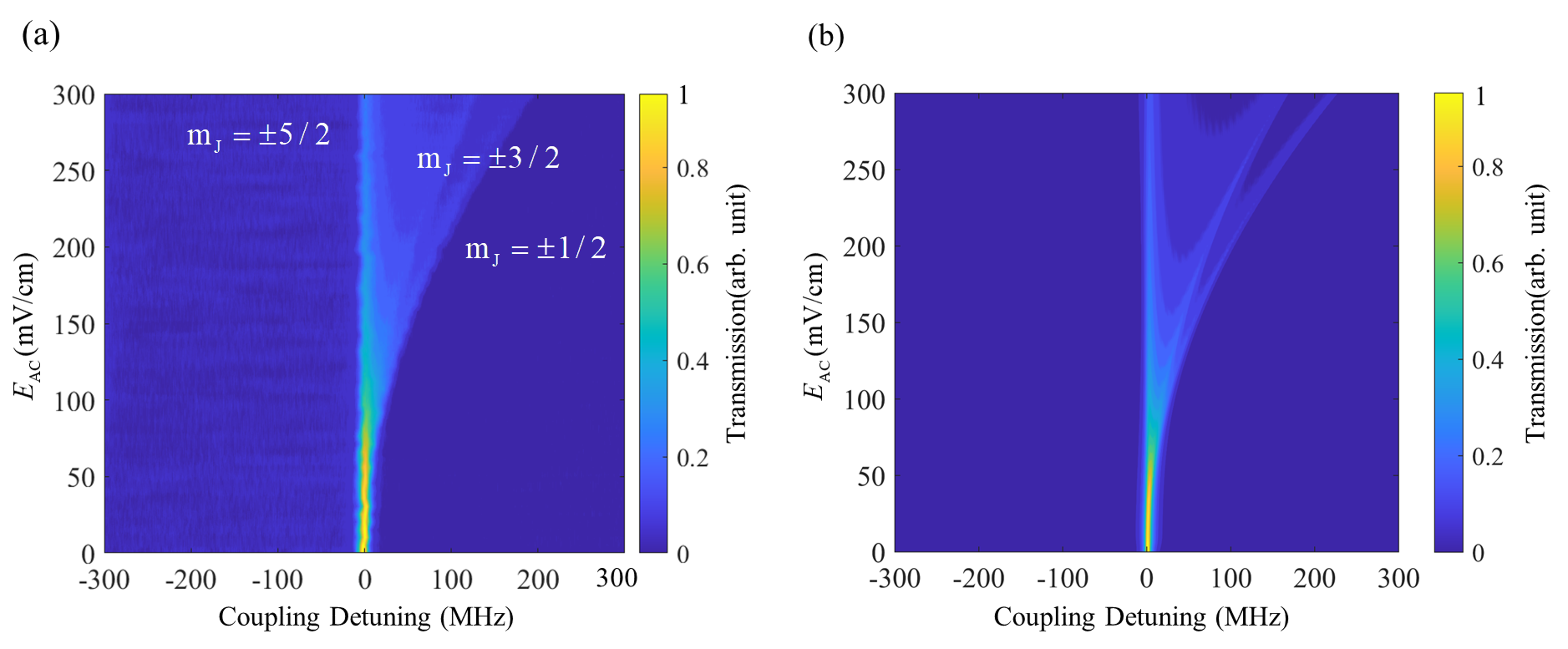} 
    \vspace{-5mm}
    \caption{The AC-Stark spectra of Rydberg atoms are measured (a) experimentally and (b) calculated theoretically.}
    \label{fig2} 
    \vspace{-3mm}
\end{figure} 

Based on the precise observation of frequency shifts induced by the combined DC and AC Stark effects in Rydberg atomic spectra, we propose a novel method to measure low-frequency electric fields as shown in Fig. \ref{fig:framework}. Unlike conventional methods, where a DC electric field typically serves only as a bias field to adjust the sensitivity or operating point of measurement, here we employ the DC electric field both as an intrinsic reference and as a coherent signal amplifier. Specifically, under the simultaneous presence of DC and AC electric fields, the originally static DC-Stark splitting of the Rydberg atomic energy levels is further modulated by the AC electric field. As a result, by cycle-averaging over the AC period, the spectra exhibit a well-defined modulation amplitude whose magnitude directly reflects the amplitude of the AC field under test.

The primary DC splitting $\Delta f_1$ is determined from a single-Lorentzian peak fit to the unmodulated EIT line. The fitted line-center shift relative to the zero-field center yields $\Delta f_1=1/2|\alpha|\,E_\mathrm{DC}^2$. With the same $E_\mathrm{DC}$ held fixed and an AC field applied, we fit the cycle-averaged Lorentz profile in Eq.~\eqref{eq:modulated_averaged_Lorentz} to the demodulated spectrum to obtain the modulation swing parameter $\Delta$, which corresponds to the secondary splitting $\Delta f_2$ induced by the combined fields and obeys $\Delta f_2=2\,|\alpha|\,E_\mathrm{AC}\,E_\mathrm{DC}$. Eliminating $E_\mathrm{DC}$ via $\Delta f_1$ then gives the AC amplitude as
\begin{equation}
E_{\mathrm{AC}}=\frac{\Delta f_2}{2\sqrt{2\,|\alpha|\,\Delta f_1}}.
\label{eq:EAC}
\end{equation}

As both $\Delta f_1$ and $\Delta f_2$ reference the same atomic polarizability $\alpha$, the result is fully traceable to intrinsic atomic properties. This removes dependence on geometric or external amplitude calibrations such as electrode spacing and, by basing the readout on frequency shifts rather than signal amplitude, substantially reduces susceptibility to laser-power fluctuations, which provides a robust, interference-resistant route to sub-MHz field metrology.

\vspace{-2mm}
\section{Experimental Setup}

% 仿真和实验设置
The experimental verification of our proposed measurement method is performed using a cesium vapor cell setup. Two counter-propagating laser beams (852 nm probe and 509 nm coupling lasers) intersect within the vapor cell, generating EIT signals that are modulated by the DC and AC electric fields. Both fields are generated by a signal generator and applied to the parallel plates inside the cell through wired connections. EIT signals are recorded and analyzed using a photodetector and an oscilloscope. Additional experimental details are provided in Appendix A \cite{appen}.

Simultaneously, numerical simulations based on the established six-level Lindblad model described earlier (see Section \ref{sys}) are carried out, matching the atomic states and coupling parameters of our experimental system. Simulation parameters are detailed in Appendix B \cite{appen}.

\begin{figure}
    \centering
    \includegraphics[width=0.8\linewidth]{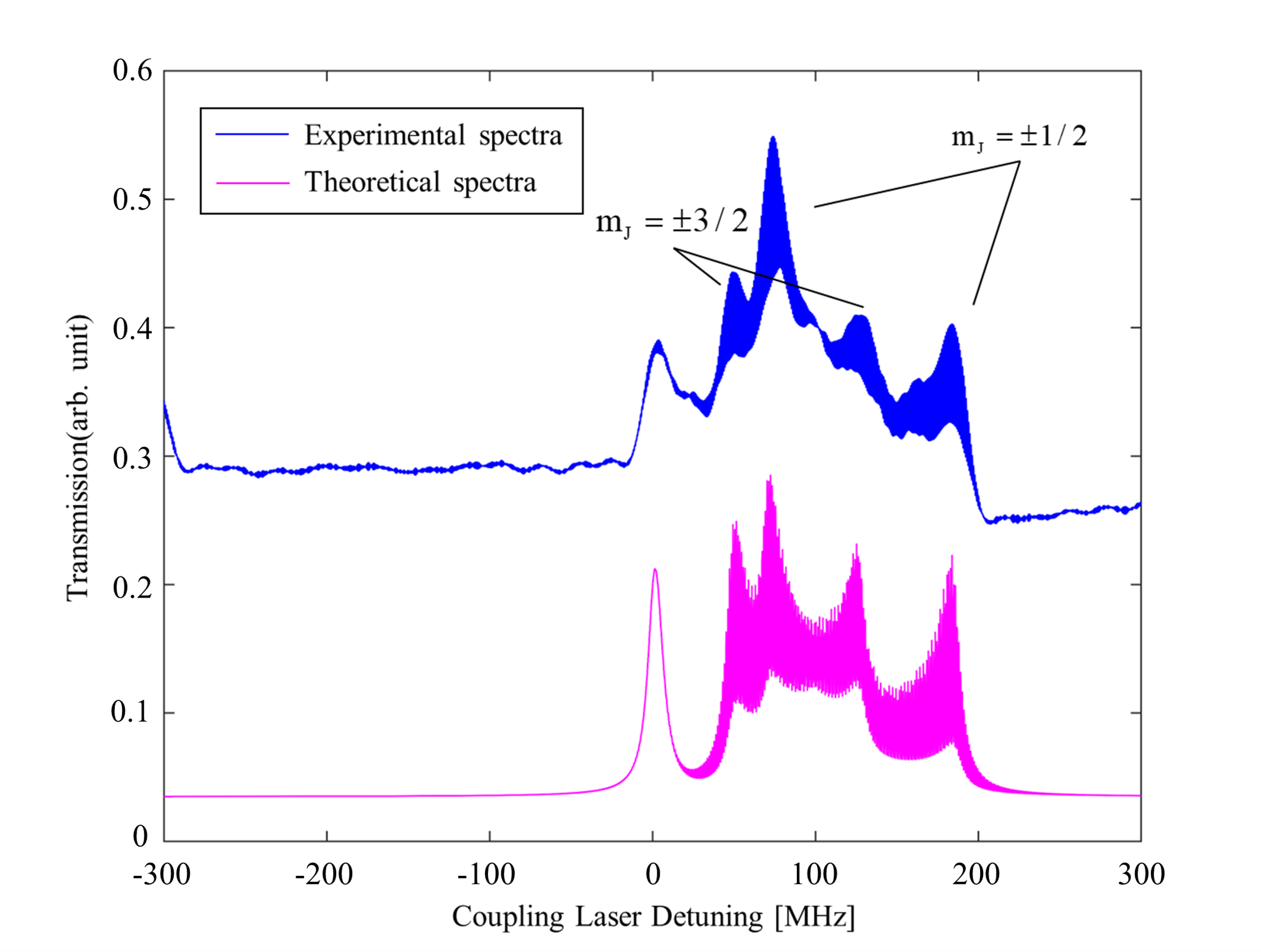} 
    \vspace{-1mm}    
    \caption{Experimental and theoretical spectra under a mixed electric field with a 220 mV/cm DC electric field and a 50 mV/cm, 30 kHz AC electric field.}
    \label{fig3} 
    \vspace{-1mm}
\end{figure}

\begin{figure}
    \centering
    \includegraphics[width=1\linewidth]{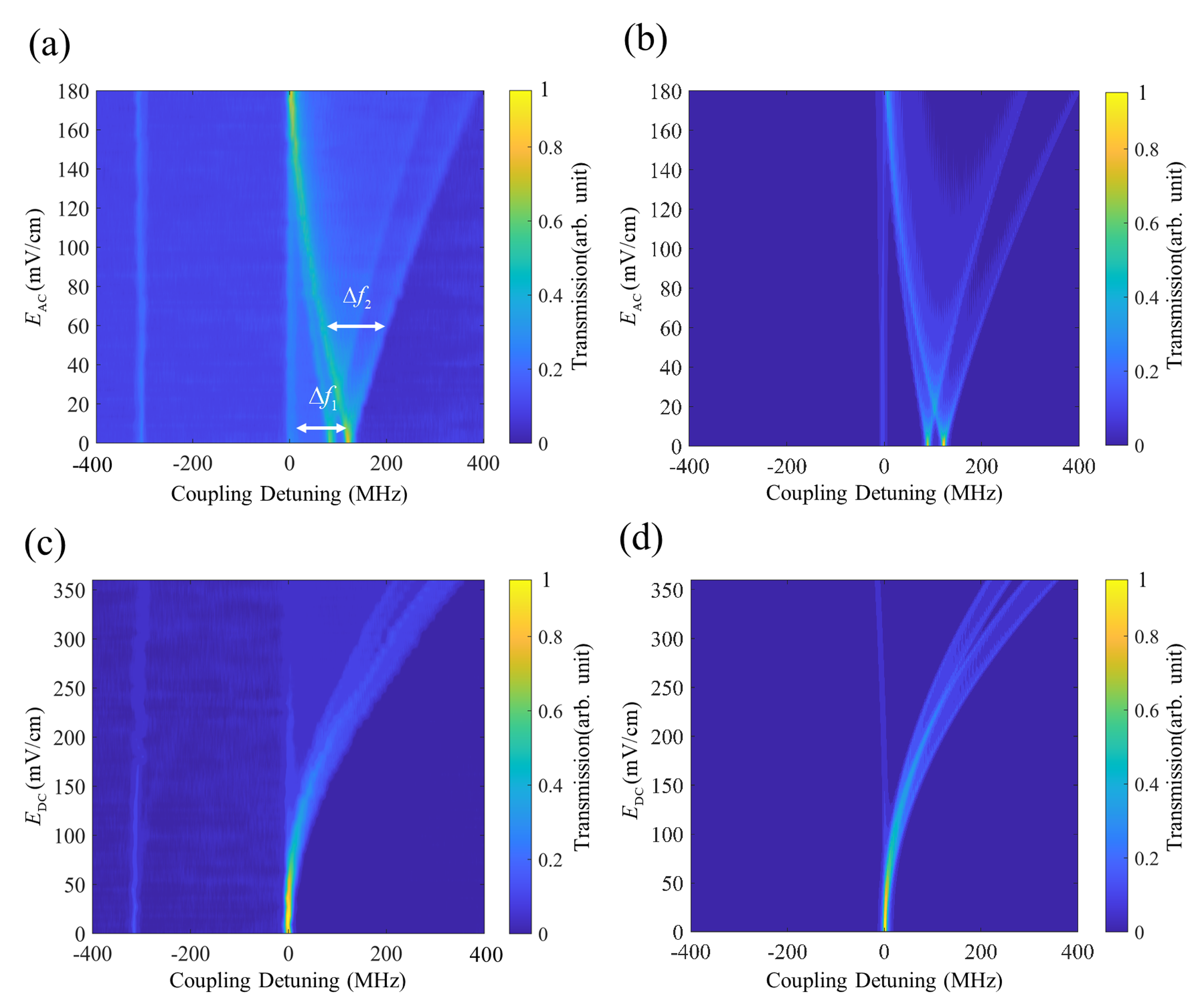} 
    \vspace{-5mm}
    \caption{Experimental (a, c) and theoretical (b, d) spectra are measured under two conditions: (a, b) with a 220 mV/cm DC field and a 0–180 mV/cm, 30 kHz AC field, and (c, d) with a 0–360 mV/cm DC field and an 18.5 mV/cm, 30 kHz AC field.}
    \label{fig4} 
    \vspace{-4mm}
\end{figure}

\vspace{-2mm}
\section{Experimental Results}

\subsection{Validation of Dual Field Response Model}

The core prediction of our model is that under a combined DC ($E_{\mathrm{DC}}$) and low-frequency AC ($E_{\mathrm{AC}}$, $f_{\mathrm{AC}}$) electric field, the time-averaged EIT spectrum exhibits characteristic splittings proportional to $E_{\mathrm{AC}} E_{\mathrm{DC}}$. Results shown in Fig. \ref{fig2}, \ref{fig3}, \ref{fig4} validate this prediction experimentally and theoretically. Fig. \ref{fig2} shows the sublevel peak splittings under different electric fields ($f_{\mathrm{AC}} = 30$ kHz) in experiment (a) and calculation (b). The observed discrepancies in frequency shifts are primarily attributed to non-uniformities in the applied electric ﬁeld, caused by imperfect alignment of the electrode plates. Since the $m_{J}=\pm1/2$ manifold has the highest polarizability, we therefore quantify limit of detection (LOD) from the splitting of that line throughout the analysis. Fig. \ref{fig3} compares experimental and theoretical spectra for a representative case ($E_{\mathrm{DC}} = 220$ mV/cm, $E_{\mathrm{AC}} = 50$ mV/cm, $f_{\mathrm{AC}} = 30$ kHz). Both show the emergence of distinct sidebands flanking the time-varying electric field-split $m_J = \pm 1/2$ and $\pm 3/2$ subpeaks. Critically, the splitting interval between these sidebands for each sublevel is directly observable. Strong qualitative agreement confirms that the model captures the essential spectral features induced by the dual-field interaction.

\begin{figure*}
\begin{center}
\includegraphics[width=1\textwidth]{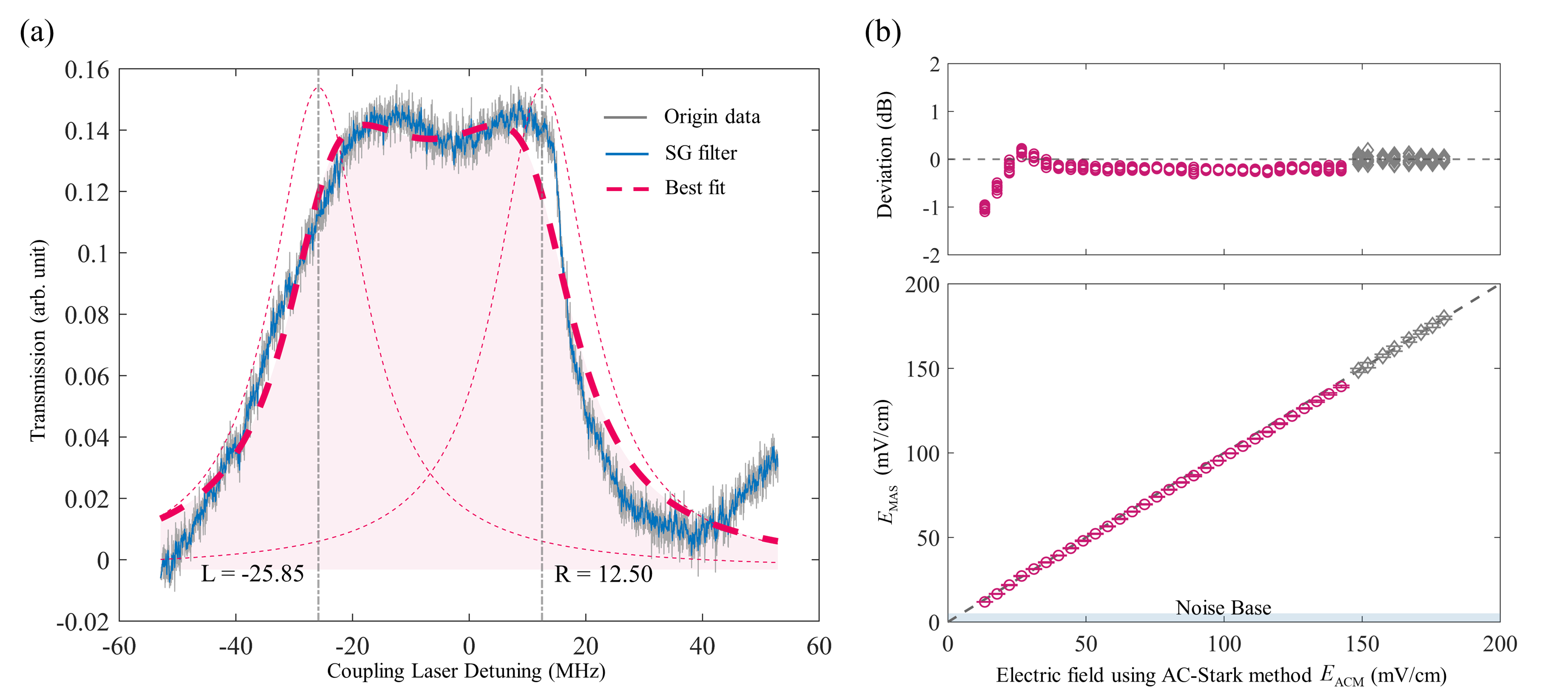} 
\end{center}
\vspace{-3mm}
\caption{(a) Representative spectrum versus coupling-laser detuning.  
    Vertical dashed lines mark $L=b-\Delta$ and $R=b+\Delta$; the shaded band highlights the integrated splitting region with $\Delta f_2=2\Delta$. 
    (b) LOD analysis of the electric field measurement. Bottom: proposed Dual Field Modulated method $E_{\mathrm{MAS}}$ (red circles) plotted against the conventional AC-Stark peak-fit result $E_{\mathrm{ACM}}$ (gray diamonds). Top: deviation expressed as $20\log_{10}(E_{\mathrm{MAS}}/E_{\mathrm{ACM}})$.}
\label{fig:fig5} % Fig.5
\vspace{-4mm}
\end{figure*}

To quantify the relationship between $\Delta f_2$ and the applied fields, we systematically varied $E_{\mathrm{AC}}$ under a fixed $E_{\mathrm{DC}}$ (220 mV/cm). Fig. \ref{fig4} experimentally (a) and theoretically (b) show that $\Delta f_2$ for the $m_J = \pm 1/2$ subpeaks increases linearly with $E_{\mathrm{AC}}$, as predicted by the model ($\Delta f_2 = |2\alpha E_{\mathrm{AC}} E_{\mathrm{DC}}|$). This linear dependence demonstrates that the DC field serves not merely as a bias, but fundamentally enables the quantification of the AC field by translating its effect into a measurable spectral splitting proportional to $E_{\mathrm{AC}}$.

Conversely, Fig. \ref{fig4} experimentally (c) and theoretically (d) show $\Delta f_2$ increasing linearly with $E_{\mathrm{DC}}$ under a fixed $E_{\mathrm{AC}}$ (18.5 mV/cm). The DC field strength thus acts as a heterodyne amplifier, enhancing the resolvability of the AC-induced splitting. However, at high $E_{\mathrm{DC}}$, experimental spectra show broadening and reduced signal-to-noise ratio compared to theory. This is attributed to increased Rydberg state coupling, electric field inhomogeneities, and stochastic processes like collisions, highlighting the need for optimizing $E_{\mathrm{DC}}$ to balance splitting magnitude and spectral clarity.

In summary, the agreement between experimental and modeled spectra, coupled with the clear demonstration of $\Delta f_2 \propto E_{\mathrm{AC}} E_{\mathrm{DC}}$, validates our quantum model for the dual-field Stark response. This forms the foundation for the calibration-free measurement method proposed in Section~\ref{sec:method}.

\vspace{-1mm}
\subsection{Performance of Method Proposed}

Fig.~\ref{fig:fig5} (a) shows a representative spectrum together with a Savitzky-Golay pre-filter (blue) and the best fit (magenta dashed) to the cycle-averaged Lorentz model in Eq. (\ref{eq:modulated_averaged_Lorentz}). The SG filter is a local-polynomial smoothing method that preserves peak height and curvature and introduces negligible phase distortion, which is therefore suitable for denoising spectra without biasing the retrieved line center or width. The nonlinear least-squares routine returns the line center $b$ and the AC-induced swing $\Delta$. The two vertical reference lines indicate $L=b-\Delta$ and $R=b+\Delta$, therefore represent the lower and upper limits of the instantaneous resonance shift during one AC cycle. The total secondary splitting is then obtained as $\Delta f_2 = 2 \Delta$. Because the waveform is a time averaged superposition of instantaneous Lorentzians centered at $b+\Delta\sin\phi$, the composite line generally exhibits shoulders. Consequently, $L$ and $R$ do not coincide with the maxima of the measured envelope; they indicate the extrema of the instantaneous center sweep, and $\Delta f_2$ is extracted from the model parameter $\Delta$, not from peak-to-peak spacing. Therefore, unlike Ref. \cite{holloway2022electromagnetically}, one cannot directly treat the shot-to-shot fluctuation of the EIT peak position as the frequency-shift uncertainty for defining the detection limit.

\begin{figure}[t!]
    \centering
    \includegraphics[width=1\linewidth]{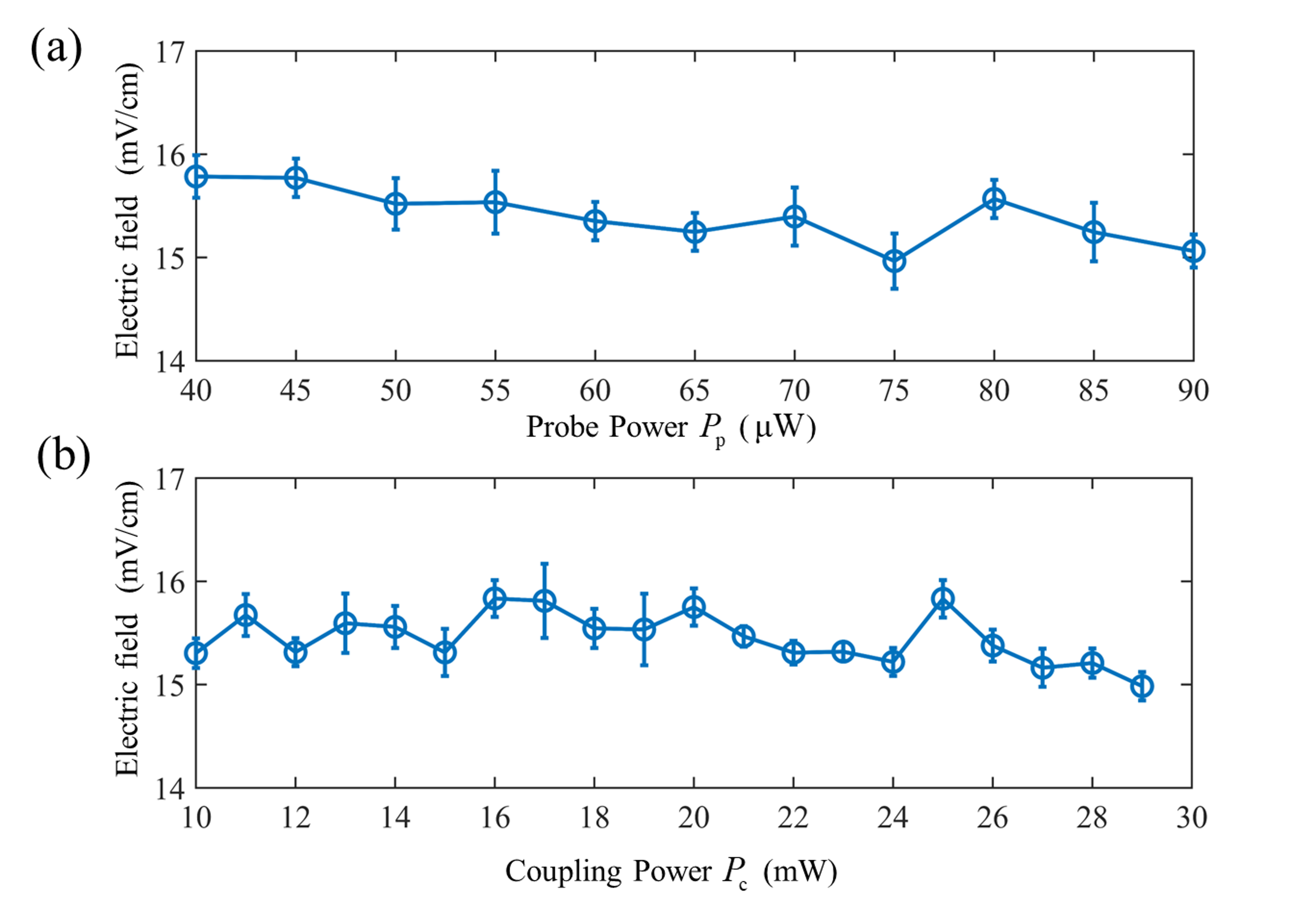} 
    \vspace{-6mm}
    \caption{Robustness to (a) probe and (b) coupling power fluctuations.}
    \label{fig:fig6} 
    \vspace{-4mm}
\end{figure}

As shown in Fig.~\ref{fig:fig5} (b), we demonstrate the LOD performance of the proposed Modulation-Average-Spectrum (MAS) readout in comparison with the conventional AC-Stark Method (ACM), which uses Stark peak fitting at the frequency of 50 MHz \cite{li2021power}. The red circles (MAS) closely match the ACM results and the deviation expressed as $20\log_{10}(E_{\mathrm{MAS}}/E_{\mathrm{ACM}})$ remains within $\pm 2$~dB (top panel), indicating negligible bias and excellent linearity. For each field, 9 single-shot measurements are performed. Error bars indicate 1$\sigma$ standard error. To quantify the detection floor, we applied the same fit model to zero-field spectra and take the 95th percentile of the pseudo-splitting distribution from 400 samples as the limit-of-blank $\Delta_{\mathrm{LOB}}$. Projecting $\Delta_{\mathrm{LOB}}$ onto the fit line yields a minimum detectable field of $E_{\min}=5.3$~mV/cm (shaded ''Noise Base'' in Fig.~\ref{fig:fig5} (b). The effective measurement bandwidth is set to $B = 520.8~\text{Hz}$, determined by the laser scan duration used in the fitting procedure. The full limit of detection calculation and statistics of the blank measurements are summarized in the Appendix C \cite{appen}.

For a fixed DC bias corresponding to a DC-only splitting of $\Delta f_1=319.7~\mathrm{MHz}$ and an AC drive at $f_{\mathrm{AC}}=30~\mathrm{kHz}$ with amplitude $E_{\mathrm{AC}}=15.4~\mathrm{mV/cm}$, Fig.~\ref{fig:fig6} evaluates robustness against optical-power fluctuations. The optical power before entering the vapor cell is measured with a photodiode power sensor (Thorlabs S121C) prior to each measurement. In panel (a), we hold the coupling power at $P_\mathrm{c}=29~\mathrm{mW}$ and sweep the probe power $P_\mathrm{p}$ from $40$ to $90~\mathrm{~\upmu W}$; in panel (b), we hold the probe power at $P_\mathrm{p}=50~\mathrm{~\upmu \mathrm{W}}$ and sweep the coupling power $P_\mathrm{c}$ from $10$ to $29~\mathrm{mW}$. Quantitatively, the fitted modulation swing $\Delta$ and the field estimate remains statistically unchanged across both sweeps: panel (a) shows a mean variation of $\pm 2.7\%$ with a point-wise root-mean-square (RMS) spread of $0.79~\mathrm{mV/cm}$ ($\approx 1.4\%$ of the mean), while panel (b) shows $\pm 2.8\%$ with a point-wise RMS spread of $0.67~\mathrm{mV/cm}$ ($\approx 1.2\%$ of the mean). Method for laser power robustness analysis is detailed in Appendix D \cite{appen}. This stability arises because $\Delta$ is tied to the instantaneous center sweep of the resonance rather than to absolute transmission, rendering the modulation-fit readout intrinsically tolerant to power noise.

\vspace{-2mm}
\section{Conclusion}

In this paper, we present a calibration-free Rydberg atomic receiving method that directly traces the sub-MHz AC field amplitudes to intrinsic atomic polarizability via two measurable Stark-induced splittings, thereby removing dependence on external calibration. An analytic, cycle-averaged Lorentz model provides a robust estimator for the AC-induced swing and, together with a Lindblad description, captures the dual-field spectral response. Experiments confirm the key prediction that the observable splitting scales linearly with the product $E_{\mathrm{AC}}E_{\mathrm{DC}}$, and the method delivers a minimum detectable field of $E_{\min}=5.3~\mathrm{mV/cm}$, while remaining insensitive to moderate optical-power fluctuations. These results establish a compact, calibration-free, and robust pathway for quantitative low-frequency sensing in domains such as underwater and subsurface communications for underwater vehicles, where antenna size and trailing-wire deployments are severely constrained. Future work will target atomic low-frequency communications, micro-cell and on-chip integration, and real-time demodulation to further extend bandwidth and stability.

\vspace{-2mm}
\bibliographystyle{IEEEtran} 
\bibliography{Bibliography}

\vfill

\end{document}